\documentclass[prd,12pt]{article}

\usepackage{diagbox}
\usepackage{tikz}
\usepackage{amsmath,amssymb,graphicx,multirow,xspace,slashed,array,booktabs}
\usepackage[colorlinks=true,urlcolor=blue,anchorcolor=blue,citecolor=blue,filecolor=blue,linkcolor=blue,menucolor=blue,pagecolor=blue]{hyperref}
\usepackage[compress,numbers]{natbib}
\usepackage{placeins}
\usepackage{subcaption}
\usepackage{booktabs}
\usepackage{blindtext, rotating}
\usepackage{afterpage}
\usepackage{enumitem}
\usepackage{marvosym}
\usepackage{authblk} 
\usepackage{verbatim}
\usepackage{soul} 
\usepackage[normalem]{ulem}
\usepackage{pifont}
\usepackage{booktabs}
\usepackage{bm}
\usepackage{cleveref}
\usepackage{siunitx}
\usepackage{tikz}
\usepackage{tikz-feynman}
\tikzfeynmanset{compat=1.1.0}
\usepackage{cases}
\usepackage{cancel}

\newcommand{\vev}[1]{ \left\langle {#1} \right\rangle }

\newcommand{\lmk}{\left(}  
\newcommand{\rmk}{\right)}

\newcommand{\bea}{\begin{array}}
\newcommand{\eea}{\end{array}}
\newcommand{\beq}{\begin{eqnarray}}
\newcommand{\eeq}{\end{eqnarray}}

\newcommand{\dd}{\mathrm{d}}

\def\SEC#1{Sec.~\ref{#1}}
\def\FIG#1{Fig.~\ref{#1}}
\def\EQ#1{Eq.~(\ref{#1})}

\usepackage{floatrow}
\newfloatcommand{capbtabbox}{table}[][\FBwidth]

\usepackage[font=footnotesize,labelfont=bf]{caption}

\usepackage{lineno}

\allowdisplaybreaks

\long\def\symbolfootnote[#1]#2{\begingroup%
\def\thefootnote{\fnsymbol{footnote}}\footnote[#1]{#2}\endgroup}

\allowdisplaybreaks

\makeatletter
	
	\@addtoreset{equation}{section}
\makeatother

\newcommand{\newc}{\newcommand}
\newc{\gsim}{\lower.7ex\hbox{$\;\stackrel{\textstyle>}{\sim}\;$}}
\newc{\lsim}{\lower.7ex\hbox{$\;\stackrel{\textstyle<}{\sim}\;$}}
\newc{\gev}{\,{\rm GeV}}
\newc{\mev}{\,{\rm MeV}}
\newc{\ev}{\,{\rm eV}}
\newc{\kev}{\,{\rm keV}}
\newc{\tev}{\,{\rm TeV}}

\def\tr{\mathop{\rm tr}}

\newc{\mz}{M_Z}
\newc{\mpl}{M_*}
\newc{\mw}{m_{\rm weak}}
\newc{\nr}[1]{N^c_R{}_{#1}}

\usepackage{accents}
\newlength{\dhatheight}

\def\beq{\begin{equation}}
\def\eeq{\end{equation}}
\def\bitem{\begin{itemize}}
\def\eitem{\end{itemize}}

\begin{document}
\baselineskip 0.6cm

\begin{titlepage}

\vspace*{-0.5cm}

\thispagestyle{empty}

\hfill KOBE-COSMO-26-08 \\

\begin{center}

\vskip 1cm

{\LARGE
Particle Production via Rippled Bubble Walls
}

\vskip 1.5cm
{Ryusuke Jinno$^{1}$, Shota Nakagawa$^{2,3}$, Yuichiro Nakai$^{2,3}$, and Yaoduo Wang$^{2,3}$}
\\*[10pt]
$^1${\it \normalsize Department of Physics, Graduate School of Science, Kobe University, \\
1-1 Rokkodai, Kobe, Hyogo, 657-8501, Japan}\\*[3pt]
$^2${\it \normalsize Tsung-Dao Lee Institute, Shanghai Jiao Tong University, \\
No.~1 Lisuo Road, Pudong New Area, Shanghai, 201210, China} \\*[3pt]
$^3${\it \normalsize School of Physics and Astronomy, Shanghai Jiao Tong University, \\
800 Dongchuan Road, Shanghai, 200240, China} \\*[3pt]

\vskip 1.0cm

\end{center}

\begin{abstract}

We investigate non-thermal particle production during first-order phase transitions in the presence of ultra-relativistic thick bubble walls with non-trivial internal structure. Extending the framework of bubble-expansion particle production, we consider bubble walls containing multiple ripples and study how such spatial modulations affect the production of heavy particles coupled to the order parameter field. By modeling an oscillatory thick-wall profile, we derive the transition probability for particle splitting processes in the wall background, and identify a new contribution associated with momentum transfer from the wall microstructure. In addition to the conventional channel, we find an enhanced production mode arising from resonant momentum exchange with the ripples. For sufficiently large numbers of ripples, the new contribution can dominate the production rate and significantly increase the abundance of particles much heavier than the phase-transition scale. Our results demonstrate that the internal structure of expanding bubble walls can play an important role in particle production and should be taken into account when assessing the cosmological implications of strongly first-order phase transitions.

\end{abstract}

\flushbottom

\end{titlepage}

\tableofcontents

\section{Introduction
\label{sec:introduction}}

First-order phase transitions in the early Universe~\cite{Witten:1984rs,Hogan:1986qda} provide a remarkably rich arena for a wide range of cosmological phenomena, such as non-thermal particle production~\cite{Watkins:1991zt,Falkowski:2012fb,Katz:2016adq,Baker:2019ndr,Chway:2019kft,Baldes:2020kam,Azatov:2020ufh,Azatov:2021ifm,Baldes:2021aph,Azatov:2022tii,Baldes:2022oev,Jinno:2022fom,Baldes:2023cih,Mansour:2023fwj,Shakya:2023kjf,Gouttenoire:2023roe,Giudice:2024tcp,Ai:2024ikj,Fujikura:2024jto,Baldes:2024wuz,An:2026sdu,Ghoshal:2026pew}, baryogenesis~\cite{Kuzmin:1985mm,Cohen:1993nk,Rubakov:1996vz,Trodden:1998ym,Riotto:1999yt,Konstandin:2011ds,Morrissey:2012db,Katz:2016adq,Fujikura:2021abj,Azatov:2021irb,Baldes:2021vyz,Fujikura:2024jto,Girmohanta:2025wcq}, primordial black holes~\cite{Sato:1981hk,Sato:1981gv,Izawa:1982cu,Hawking:1982ga,Kodama:1982sf,Moss:1994iq,Khlopov:1998nm,Jedamzik:1999am}, primordial magnetic fields~\cite{Vachaspati:1991nm,Baym:1995fk,Sigl:1996dm,Ahonen:1997wh}, and stochastic gravitational-wave (GW) backgrounds~\cite{Kosowsky:1991ua,Kosowsky:1992rz,Kosowsky:1992vn,Kamionkowski:1993fg} (for reviews, see e.g. Refs.~\cite{Weir:2017wfa,Mazumdar:2018dfl,Caprini:2019egz,Hindmarsh:2020hop,Athron:2023xlk,Caprini:2024hue}). Of particular interest are supercooled phase transitions, where the Universe remains trapped in a metastable vacuum until temperatures far below the critical temperature. The large vacuum energy released during such phase transitions can drive ultra-relativistic bubble expansion and trigger highly non-equilibrium dynamics, leading to distinctive cosmological signatures. Supercooled transitions can generate non-thermal particle populations through several physically distinct mechanisms. Bubble collisions and non-equilibrium order-parameter dynamics can directly create coupled particles~\cite{Watkins:1991zt,Falkowski:2012fb,Katz:2016adq,Mansour:2023fwj,Shakya:2023kjf,Giudice:2024tcp,An:2026sdu,Ghoshal:2026pew}; Plasma particles can scatter from bubble walls and undergo splittings~\cite{Azatov:2020ufh,Azatov:2021ifm,Baldes:2022oev,Ai:2024ikj}, a process that has been found to play a central role in determining ultra-relativistic bubble walls~\cite{Bodeker:2017cim,Gouttenoire:2021kjv,Long:2024sqg}; Phase-dependent reflection and transmission can filter thermal populations into non-thermal ones~\cite{Baker:2019ndr,Chway:2019kft,Jinno:2022fom}; Particles produced or accelerated at the wall can form dense shells whose energetic collisions generate heavier states~\cite{Baldes:2023cih}, although these shells do not generally free-stream~\cite{Baldes:2024wuz}; In confining theories, wall crossing can stretch strings that fragment into composite states~\cite{Baldes:2020kam,Baldes:2021aph}, while bound-state formation can further modify their abundance~\cite{Gouttenoire:2023roe}. Among these, the authors of Ref.~\cite{Azatov:2020ufh,Azatov:2021ifm} pointed out that particle splitting could serve as a novel mechanism for non-thermal particle production.

In the wall frame, particles from the plasma are Lorentz boosted to energies of $\sim \gamma_{w} T$, where $\gamma_{w}$ represents the wall Lorentz factor, enabling the production of particles with masses far exceeding the ambient plasma temperature $T$. While a smooth and monotonic bubble-wall profile adopted in these analyses provides a useful starting point, the scalar-field configuration can be appreciably more complex in supercooled phase transitions. When the scalar field tunnels out of the false vacuum, the tunneling endpoint generally lies far from the true-vacuum minimum. Consequently, the field does not settle immediately into the vacuum configuration but instead overshoots the minimum and undergoes coherent oscillations around it. As the bubble expands, these oscillations can propagate through the bubble interior and imprint themselves on the wall configuration. The resulting wall profile is therefore not necessarily described by a single smooth interpolation between the false and true vacua. Instead, it may contain a sequence of oscillatory features or ripples associated with the coherent motion of the scalar field. Since the non-thermal particle production is governed by the momentum transfer provided by the wall background, the presence of these oscillatory structures can significantly modify the production process. Understanding their impact is therefore essential for obtaining a realistic description of particle production in supercooled phase transitions.

In the present paper, we investigate non-thermal particle production from relativistic thick bubble walls containing multiple ripples. We model the oscillatory wall profile by a simple periodic ansatz and compute the probability for heavy particle production. It is revealed that the ripple structure generates new momentum-transfer channels that are absent for a smooth wall. In addition to the usual contribution present in smooth-wall profiles, we then identify a resonant contribution originating from the characteristic wavelength of the ripples. This new channel can noticeably enhance the production rate of heavy particles. Our results demonstrate that the internal structure of bubble walls can play an important role in non-thermal particle production and should therefore be taken into account in realistic studies of supercooled phase transitions.

The rest of the paper is organized as follows. Sec.~\ref{splitting} introduces a simple model of a relativistic bubble wall with an oscillatory internal structure and derives the probability for non-thermal particle production in the wall background. We identify both the conventional momentum-transfer channel present in smooth-wall profiles and a new resonant channel induced by the ripple structure. In Sec.~\ref{pheno}, we explore the phenomenological implications of these results, including the friction on the bubble walls from the produced particles and the resulting relic abundance. In \SEC{sec:evolution}, we also discuss the effective number of the ripple in realistic bubble profiles and present an example based on a classical scale invariant model. Sec.~\ref{conclusion} is devoted to discussion and conclusions.

\section{Particle splitting via rippled walls}
\label{splitting}

We consider the splitting of an incident particle,
\begin{equation}
    h + {\rm (wall~background)} \to \phi + \phi \ ,
\end{equation}
in the background of a relativistic bubble wall containing multiple ripples.
Such ripples of the wall are commonly expected in supercooled phase transitions.
As we will show, the ripple structure gives rise to a new resonantly enhanced contribution in addition to the conventional momentum-transfer channel identified in Refs.~\cite{Azatov:2020ufh,Azatov:2021ifm}.

\subsection{Setup
\label{sec:setup}}
 
\begin{figure}
    \centering
    \includegraphics[width=0.75\linewidth]{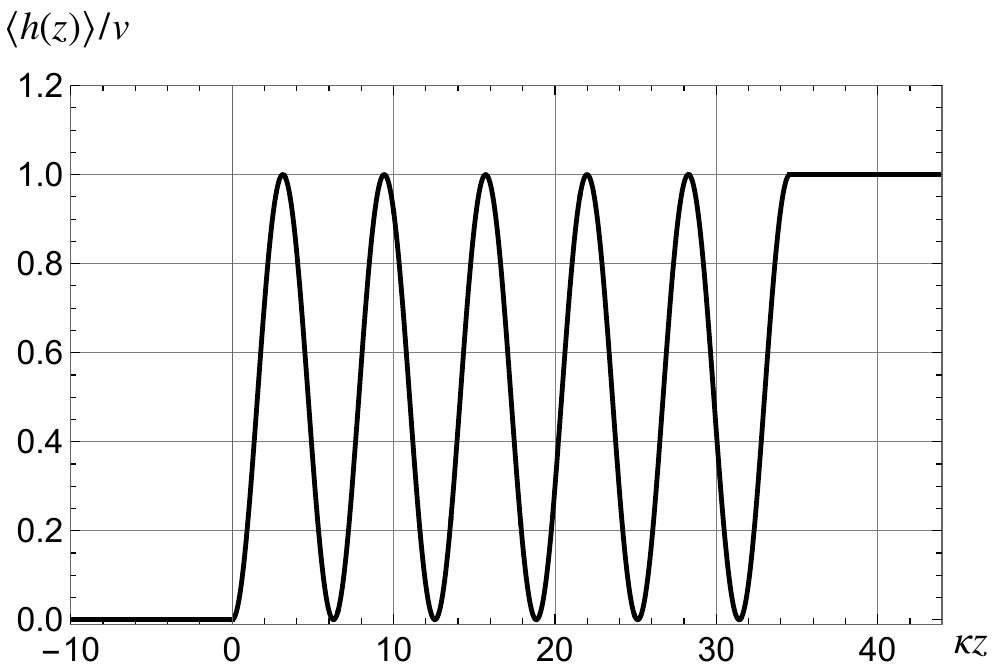}
    \vspace{3mm}
    \caption{
    A modeling of the bubble wall profile $\langle h(z) \rangle$ in Eq.~\eqref{bubble-profile} for $n=5$.
    }
    \label{fig:profile}
\end{figure}

To simplify our analysis,  let us introduce a toy model (presented in Ref.~\cite{Azatov:2021ifm}) with the Lagrangian density of two scalar fields,
\begin{equation}
    \mathcal{L} = |\partial H|^2 + \frac12 (\partial \phi)^2 - V(H, \phi) \ ,
\end{equation}
where
\begin{equation}
    V(H, \phi) = V_{\rm eff}(H) +  \frac {M_\phi^2}2 \phi^2 + \frac \lambda2 \phi^2 |H|^2 \ .
\end{equation}
Here, $\phi$ denotes a real scalar field with a mass parameter $M_\phi$, $H$ is a complex scalar field coupled to $\phi$ with a coupling constant $\lambda$, and $V_{\rm eff}(H)$ is its effective potential.
We consider the production of heavy $\phi$ as a particle that could potentially be a dark matter (DM).
With this in mind, we impose a $Z_2$ symmetry under which $\phi$ is odd to guarantee its stability.
We assume that $V_{\rm eff}(H)$ causes a supercooled first-order phase transition in the early universe.
The nucleation of bubbles completes through the tunneling process of $h\equiv \sqrt{2}|H|$, and the exit point tends to be far from the true vacuum in this case.
The subsequent oscillation of $h$ can leave a lot of ripples on the bubble-wall profile.
A realistic bubble-wall profile depends on the evolution of the scalar field $h$, which will be discussed in \SEC{sec:evolution} for specific forms of the effective potential $V_{\rm eff}(H)$. For now, we parameterize the bubble-wall profile in the wall rest frame as
\begin{equation}
    \langle h(z) \rangle = \begin{cases}
        v \ ,& \displaystyle z \ge \frac{(2n+1)\pi}{\kappa}, \\[0.3cm]
        \displaystyle v \times \frac{1-\cos( \kappa z ) }{2} \  ,& \displaystyle 0 \le z \le \frac{(2n+1)\pi}{\kappa} \ , \\[0.3cm]
        0 \ , & z<0 \ .
    \end{cases}
    \label{bubble-profile}
\end{equation}
Here the frame is set so that the ambient plasma propagates in the positive $z$-direction, and we assume that the bubble wall is locally planar, which is justified because the bubble radius is much larger than the microscopic length scale relevant to the particle production process.
We define the vacuum expectation value (VEV) at the true minimum as $v$, and $n$ is the number of ripples on the wall with the length scale $\kappa^{-1}$ (see \FIG{fig:profile}).

The presence of the bubble background alters the nature of $\phi$ particles between the phases, such as mass and interaction vertex. 
When the nature changes non-adiabatically, the particle production can occur through the bubble walls \cite{Bodeker:2017cim,Azatov:2020ufh,Azatov:2021ifm,Azatov:2021irb,Azatov:2022tii,Azatov:2024crd}.
In our setup, the relevant process is $h\to \phi\phi$, where an $h$ particle in the thermal bath splits into two $\phi$ particles across the wall.
Since the wall background is invariant under time translations and translations parallel to the wall, energy and transverse momentum are conserved, whereas the wall can absorb or supply momentum along the $z$ direction. Together with the large Lorentz boost of the incoming $h$ particle in the wall rest frame, this allows the process $h\to\phi\phi$ to occur even when $\phi$ is much heavier than $h$.

The incident energy is dominated by the largely boosted momentum, $E\sim p_z \sim \gamma_w T_{n}$, where $\gamma_w\equiv 1/\sqrt{1-v_w^2}$ is the gamma factor of the wall with $v_w$ the wall velocity, and $T_n$ is the nucleation temperature, which is the typical temperature scale for the production process.
Taking a small boost along the transverse momentum,
we define the following kinematics for $h$ and two $\phi$ particles:
\begin{align}
    p &= \left(E, 0, \sqrt{E^2-m^2(z)} \right), \nonumber\\[1ex]
    k_1 &= \left(xE, -\vec k_\perp, \sqrt{x^2E^2-|\vec k_\perp|^2-M^2(z)} \right),\\[1ex]
    k_2 &= \left((1-x)E, \vec k_\perp, \sqrt{(1-x)^2E^2-|\vec k_\perp|^2-M^2(z)} \right),\nonumber
\end{align}
where $x$ is the energy fraction taken by the first $\phi$ particle, 
and the effective masses are given by  
\begin{align}
    m^2(z) &= \left. \dfrac{\partial^2 V(H+\langle H \rangle, \phi )}{\partial H \partial H^\dagger} \right|_{H, \phi = 0} \ ,\\[1ex]
     M^2(z) &= \left. \dfrac{\partial^2 V(H+\langle H \rangle, \phi )}{\partial \phi^2} \right|_{H, \phi = 0} = M_\phi^2 + \frac{1}{2}\lambda \langle h(z) \rangle^2 \ .
\end{align}
Although the masses change as the particles go through the walls, we assume that $M_\phi\gg \lambda |\langle h\rangle|$ and $m(z)$ is negligible compared to the typical energy scale, so that $m(z)\simeq 0$ and $M(z)\simeq M_\phi$.
This large hierarchy results in the Boltzmann suppression of the thermal abundance of $\phi$. 
Rather than the $z$-dependence of mass, the alternation of the vertex contributes to the production of $\phi$ in this setup.
According to our assumed profile in \EQ{bubble-profile}, the interaction Hamiltonian gives the corresponding vertex in the form of
\begin{equation}
    V(z) = \begin{cases}
        V_h \equiv  \lambda v \ ,& \displaystyle z \ge  \frac{(2n+1)\pi}{\kappa} \ , \\[0.3cm]
        \displaystyle V_{w} \equiv V_h \times \frac{1-\cos( \kappa z ) }{2} \ ,&  \displaystyle 0 \le z\ < \frac{(2n+1)\pi}{\kappa} \ , \\[0.3cm]
         V_s \equiv 0 \ ,& z<0 \ .
    \end{cases}
    \label{vertex}
\end{equation}

\subsection{Transition splitting}

We estimate the probability that an $h$ particle splits into two $\phi$ particles in the presence of the bubble-wall profile (\ref{bubble-profile}).
At leading order in the interaction, evaluating the field contractions and integrating over the final-state phase space, one finds the expression of the splitting probability as
\begin{equation}
P_{h\to \phi\phi} = \dfrac{1}{2} \times \dfrac{1}{2E}  \int \frac{d^3 k_1d^3k_2}{(2\pi)^6 2k_1^02k_2^0} 
(2\pi)^3 \delta^2 \lmk \vec{p}_\perp - \vec k_{1\perp} - \vec k_{2\perp}\rmk 
\delta \lmk E - k_1^0 -k_2^0\rmk
\left\vert \mathcal M  \right\vert^2,
\label{probability}
\end{equation}
where $\mathcal{M}$ is the invariant matrix element {and the prefactor $1/2$ is due to the phase space of two identical outgoing $\phi$ particles}.

Since the solution to the Klein-Gordon equation is $\sim e^{-iEt+ip_\perp\cdot x_\perp}\chi(z)$, one can see the conservation of energy and the transverse momentum but no momentum conservation in the $z$-direction.
The wave functions of both $h$ and $\phi$ excitation modes deviate from plane wave solutions due to the presence of the classical background $\langle h(z) \rangle$.
For $\kappa \ll p_z, k_{1z}, k_{2z}$ with $p_z \simeq \gamma_w T_n$, the bubble-wall profile varies on a length scale much larger than that of those wave functions. The particles hence experience the wall background as a slowly varying medium, allowing the use of the WKB approximation. 
The wave function can be approximated as
\begin{equation}
    \chi_p(z) \sim \sqrt{\frac{p_{z}(z=0)}{p_z(z)}} \exp\left( i \int_0^z p_z(z'){d} z' \right) \approx e^{i p_z z} \ ,
    \label{WKB}
\end{equation}
where we ignore the $z$ dependence of $p_z$ as mentioned in \SEC{sec:setup}.
Then, the scattering amplitude is given by
\begin{equation}
    \mathcal{M} = \int_{-\infty}^\infty { d}  z \,  V(z) \chi_p(z)\chi_{k_1}^*(z)\chi^*_{k_2}(z) \approx \int_{-\infty}^\infty { d}  z \, V(z) e^{i \Delta p_z z} \ ,
    \label{matrix_element}
\end{equation}
where the momentum transfer is defined by $\Delta p_z = p_z - k_{1z} - k_{2z}$.

\subsubsection{Scattering amplitude}

Let us start with the estimate of the scattering amplitude.
Using the interaction vertex in \EQ{vertex}, we obtain the matrix element as
\begin{align}
     \mathcal{M} 
     &=\int_{-\infty}^0 { d}  z \, V_s e^{i \Delta p_z z} 
     + \int_0^{(2n+1)\pi / \kappa} {d}  z \, V_{w}(z) e^{i \Delta p_z z}
     + \int_{(2n+1)\pi / \kappa}^\infty { d}  z \, V_h e^{i \Delta p_z z}\nonumber\\[1ex]
     &= 0 - iV_h \frac{\kappa^2+ e^{(2n+1)i\pi  \Delta p_z / \kappa}(2\Delta p_z^2 - \kappa^2)}{2\Delta p_z (\Delta p_z -\kappa) (\Delta p_z + \kappa)} + \frac{V_h}{i \Delta p_z} (-e^{(2n+1)i\pi  \Delta p_z/\kappa} + e^{i\infty})\nonumber\\[1ex]
     &= -\frac{i\kappa^2 V_h}{2} \dfrac{1+e^{(2n+1)i\pi  \Delta p_z/\kappa} }{ \Delta p_z ( \Delta p_z-\kappa) ( \Delta p_z+\kappa)} \ .
\end{align}
Note that we have abandoned the boundary term $\sim  e^{i\infty}$ in the final expression, which follows the implicit $i\epsilon$ prescription~\cite{Bodeker:2017cim,Azatov:2021ifm}.
The squared amplitude $|\mathcal{M}|^2$ splits into parts with different pole structures, denoted respectively by $ |\mathcal{M}|^2_{1-5}$, as follows:
\begin{equation}
    |\mathcal{M}|^2 =\sum_{i=1}^5 |\mathcal{M}|^2_i\ ,
\end{equation}
where
\begin{align}
    |\mathcal{M}|^2_1 &=  \dfrac{ V_h^2}{\Delta p_z^2}\cos^2\left[\dfrac{(2n+1)\pi \Delta p_z }{2\kappa}\right] \ , \\[1ex]
      |\mathcal{M}|^2_2 &=  \dfrac{ V_h^2}{4(\Delta p_z -\kappa)^2}\cos^2\left[\dfrac{(2n+1)\pi \Delta p_z }{2\kappa}\right] \ , \\[1ex]
        |\mathcal{M}|^2_3 &=  -\dfrac{3 V_h^2}{4\kappa(\Delta p_z -\kappa)}\cos^2\left[\dfrac{(2n+1)\pi \Delta p_z }{2\kappa}\right] \ , \\[1ex]
          |\mathcal{M}|^2_4 &= \dfrac{ V_h^2}{4(\Delta p_z +\kappa)^2}\cos^2\left[\dfrac{(2n+1)\pi \Delta p_z }{2\kappa}\right] \ , \\[1ex]
            |\mathcal{M}|^2_5 &= \dfrac{3 V_h^2}{4\kappa(\Delta p_z +\kappa)}\cos^2\left[\dfrac{(2n+1)\pi \Delta p_z }{2\kappa}\right] \ .
\end{align}

\subsubsection{Probability}

In the following calculations, we assume $E\gg M_\phi$ to derive the analytical formula.
In this case, the momentum transfer can be approximated by
\begin{align}
\Delta p_z &= E -\sqrt{x^2E^2 -\vec k_{\perp}^2-M_\phi^2} -\sqrt{(1-x)^2E^2 -\vec k_{\perp}^2-M_\phi^2}\nonumber\\
&\approx \frac{M_\phi^2 + \vec k_\perp^2}{2x(1-x)E} \ .
\label{transfer}
\end{align}
The phase space of the outgoing $\phi$ particles can be reduced as
\begin{equation}
   d^3k_{1,2}=2\pi \dfrac{k_{1,2}^0 k_{1,2 \perp}}{k_{1,2z}} d k_{1,2\perp} d k^0_{1,2} \approx \pi d k^2_{1,2\perp} d k^0_{1,2} \ ,
\end{equation}
with $k_{1,2\perp}\equiv |\vec k_{1,2\perp}|$.
Thus, the splitting probability (\ref{probability}) is simplified to
\begin{align}
 P_{h\to\phi\phi} \simeq \dfrac{1}{2} \int_0^1 d x \int d k_\perp^2  \,  \dfrac{ \Theta(E-2M_\phi)}{64\pi^2 x(1-x) E^2} |\mathcal{M}|^2 \ ,
\end{align}
where $\Theta(E-2M_\phi)$ denotes the step function.

We now estimate each contribution of $|\mathcal{M}|^2$ to the splitting probability.
Firstly, $|\mathcal{M}|^2_3$ has a subdominant contribution to the splitting probability because its leading part is asymmetric around $\Delta p_z = \kappa$ and cancels upon integration over the resonance region.\footnote{This contribution can be effective in the adiabatic regime (see Eq.~(\ref{non-adiabatic})), $E \lesssim 2M_\phi^2/\kappa$, corresponding to the pole $\Delta p_z\approx \kappa$.
Since we are interested in the non-adiabatic regime, we have numerically confirmed that  $|\mathcal{M}|^2_3$  contribution is subdominant and does not affect our conclusion.
}
In addition, $|\mathcal{M}|^2_{4,5}$ do not contribute for $\Delta p_z>0$.
Thus we evaluate the probability for $|\mathcal{M}|^2_1$ and $|\mathcal{M}|^2_2$.
The contribution from $|\mathcal{M}|_1^2$ is expressed as
    \begin{align}
    P_{h\to \phi\phi}^{\Delta p_z \approx 0} &\simeq \dfrac{1}{2}\int_0^1 d x \int d k_\perp^2 x(1-x) \frac{V_h^2 \Theta(E-2M_\phi)}{16\pi^2(M_\phi^2+k_\perp^2)^2}\cos^2\left[\dfrac{(2n+1) \pi (M_\phi^2+k^2_{\perp})}{4x(1-x)E\kappa}\right] \nonumber \\[1ex]
    &\simeq \dfrac{1}{2}\int_0^1 d x\, x(1-x)\dfrac{V_h^2  \Theta(E-2M_\phi)}{16\pi^2 M_\phi^2} A(x) \ ,
    \label{pole0}
\end{align}
where an auxiliary function $A(x)$, defined as 
    \begin{align}
         A(x) = 
        \cos ^2\left[\frac{(2 n+1) \pi  M_\phi^2}{4  x (1-x) E \kappa }\right]
          +\frac{(2n+1) \pi  M_\phi^2}{4x(1-x) E \kappa} \lmk  \int_0^{\frac{(2 n+1) \pi  M_\phi^2}{2  x (1-x)E \kappa }} d t \, \dfrac{\sin t}{t} -\frac{\pi}{2}\rmk ,
    \end{align}
is introduced.

\begin{figure}
    \centering
    \includegraphics[width=0.8\linewidth]{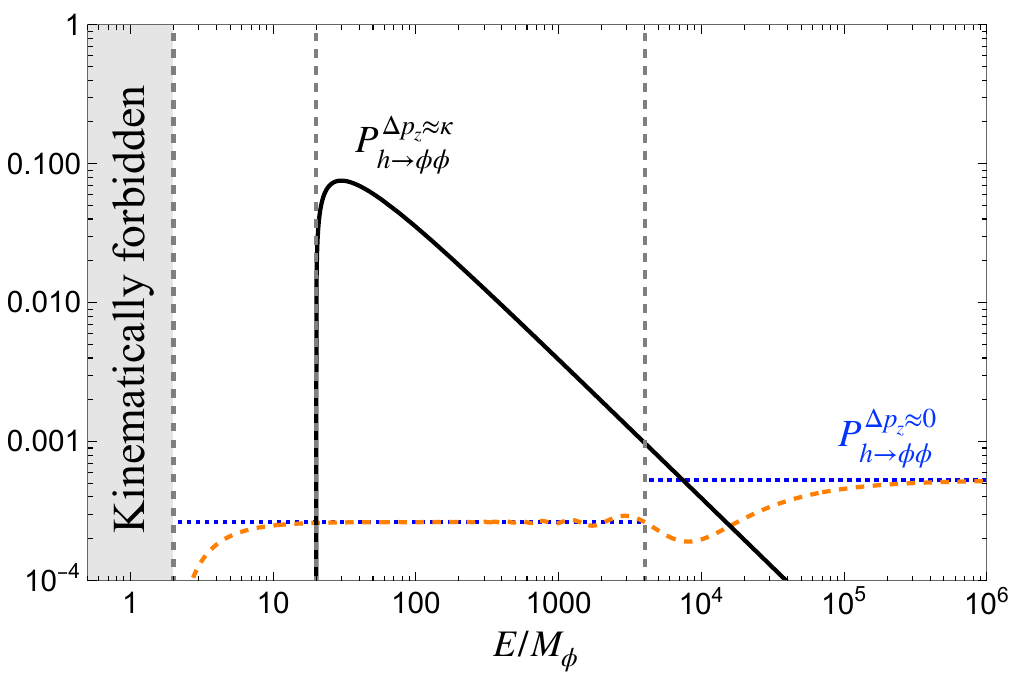}
    \caption{
    The splitting probability as a function of $E/M_\phi$ for a parameter set, $n=100$, $\kappa=M_\phi/10$, and $V_h=M_\phi$.
    The black solid and blue dotted lines show the contributions of $\Delta p_z=\kappa$ and $\Delta p_z=0$, respectively.
    From the left, the gray dashed lines correspond to the thresholds indicated in the analytic functions, (\ref{lowE_approx}), (\ref{prob_pzk}), and (\ref{highE_approx}).
    The orange dashed line represents the numerically estimated contribution of $\Delta p_z=0$ (\ref{pole0}).
    }
    \label{fig:prob}
\end{figure}

Let us find the asymptotic forms of the probability given in \EQ{pole0}.
The function $A(x)$ is controlled by $(2n+1)\Delta p_z/\kappa \simeq nM_\phi^2/[x(1-x)E\kappa]$ for $n\gg1$.
In the high-energy limit, $E\gg nM^2_\phi/\kappa$, as $A(x)\to1$, we obtain
\begin{align}
    P_{h\to \phi\phi}^{\Delta p_z \approx 0}\approx \dfrac{V_h^2}{192 \pi^2 M_\phi^2} 
    \Theta\lmk E - \dfrac{(4n+2) M_\phi^2}{\kappa}\rmk.
    \label{highE_approx}
\end{align}
The description by the step function is based on the approximation, 
\begin{equation}
    \int_0^1 d x \, 6x(1-x) \cos^2 \left[ \dfrac{a \pi}{4x(1-x)} \right] \approx \dfrac{1+\Theta(1-a)}{2} \ , \quad \max(a, a^{-1})\gg 1 \ .
\end{equation}
On the other hand, in the low-energy limit, $2M_\phi\ll E \ll nM_\phi^2/\kappa$, $A(x)\to1/2$, and the probability is reduced by half as
\begin{align}
    P_{h\to \phi\phi}^{\Delta p_z \approx 0}\approx \dfrac{V_h^2}{384 \pi^2 M_\phi^2} 
    \Theta\lmk E -2M_\phi\rmk.
    \label{lowE_approx}
\end{align}
Note that our result for $n=0$ and $\Delta p_z \ll \kappa$ is consistent with that of the previous work \cite{Azatov:2021ifm} up to the prefactor.\footnote{
We find a factor of $1/8$ difference in the matrix elements (2.18) compared to Ref.~\cite{Azatov:2021ifm}.
}
In the case of $n=0$ or Ref.~\cite{Azatov:2021ifm}, the contribution (\ref{lowE_approx}) in the low-energy range can be ignored. 
However, this contribution can be seen for $n\gg1$.
In \FIG{fig:prob}, we show the results of the probability for $n=100$, $V_h=M_\phi$, and $\kappa=M_\phi/10$.
The blue dotted lines represent the analytical results, \EQ{highE_approx} and \EQ{lowE_approx}.
The range of $E<2M_\phi$ is kinematically forbidden as shown in (\ref{lowE_approx}).
If the typical energy scale is lower than the threshold, $\gamma_w T_n < (2n+1)M_\phi^2/\kappa$ for $n\gg1$, the contribution of \EQ{lowE_approx} should be taken into account rather than \EQ{highE_approx}, which will be discussed in the next section. 
The orange dashed line represents the numerically estimated contribution (\ref{pole0}), which is consistent with the analytical results except for the transition ranges (represented by the gray dashed lines).

Next, we study the contribution from $|\mathcal{M}|^2_2$ at the pole of $\Delta p_z = \kappa$, which exhibits the characteristic effect of the finite bubble wall width.
Taking the number of ripples as $n\gg1$ and noting that
\begin{equation}
    \int_{-\infty}^\infty { d} x \frac{\sin^2 a x}{a \pi x^2} = 1 \ ,
\end{equation}
we obtain
\begin{equation}
      |\mathcal{M}|^2_2 \approx  \dfrac{ (2n+1)\pi^2 V_h^2}{8\kappa}\delta(\Delta p_z-\kappa) \ .
\end{equation}
Thus, the probability reads
\begin{align}
      P^{\Delta p_z \approx \kappa}_{h\to \phi\phi} &\simeq \dfrac{1}{2} \dfrac{(2n+1)V_h^2}{512 E \kappa} \int_0^1 dx \int dk_\perp^2 \frac{\Theta(E-2M_\phi)}{x(1-x)E}\delta\lmk\frac{M_\phi^2+k_\perp^2}{2x(1-x)E}-\kappa\rmk \nonumber \\[1ex]
       &= \dfrac{1}{2} \dfrac{(2n+1)V_h^2}{256 E \kappa} \Theta(E-2M_\phi) 
       \int_0^1 d x \, \Theta \lmk  \kappa -  \dfrac{M_\phi^2}{2x(1-x)E} \rmk \nonumber \\[1ex]
       &=  \dfrac{(2n+1)V_h^2}{512 E \kappa} 
       \sqrt{1-\dfrac{2M_\phi^2}{E \kappa}} \Theta\lmk E-\dfrac{2M_\phi^2}{\kappa}\rmk \Theta(E-2M_\phi) \ .
       \label{prob_pzk}
\end{align}
In the first line, we have used \EQ{transfer}. 
The contribution $P^{\Delta p_z \approx \kappa}_{h\to \phi\phi}$ can take dominance for $E\gtrsim 2M_\phi^2/\kappa$ and $n\gg1$ compared to $P_{h\to\phi\phi}^{\Delta p_z\approx 0}$, because of the prefactor $(2n+1)$,\footnote{\EQ{prob_pzk} shows unitarity violation for extremely large $n$. 
Although we decompose the squared matrix element into five contributions $|\mathcal{M}|^2_i$, we have confirmed that it does not break unitarity by numerically estimating the probability with the full amplitude $|\mathcal{M}|^2$.
As larger $n$ means effectively larger coupling, this can be attributed to the poor perturbativity.
Thus we focus on the range of $n$ where unitarity is conserved, $P_{h\to\phi\phi}<1$.}
as shown in \FIG{fig:prob} for $n=100$.

\subsection{Physical interpretation
\label{sec:interpretation}}

We have obtained the dependence of the splitting probability on the incident energy of $h$.
Here let us discuss how the multiple ripples induce the enhancement of the scattering rate.

Through the wall, the $h$ particles undergo the oscillation of the interaction strength $V(z)$ in \EQ{vertex} repeatedly.
If the oscillation frequency $\kappa$ is very small, the $h$ particle state remains in the initial state and any excited mode does not appear. 
If the frequency rapidly changes, the particle state cannot follow the temporal state, and excited states are induced.
This phenomena is well-known in various physics situations, such as neutrino flavor conversion in matter and particle production in time-dependent backgrounds, and is understood by the non-adiabaticity of the system.
In our wall's rest frame, the time scale of the particle state is given by the variation of energy in the $z$-direction, $\Delta p_z^{-1}$.
If the system is highly non-adiabatic, in other words, 
\beq
\Delta p_z < \kappa \ ,
\label{non-adiabatic}
\eeq 
the particle production occurs efficiently.

Due to the multiple non-adiabatic changes of the interaction strength, the splitting probability is resonantly enhanced.
The invariant matrix element inside the wall can be expressed as
\begin{align}
    \mathcal{M}^{(n)} 
    &= \int_0^{nL} d z \, e^{i \Delta p_z z} V_w(z) \nonumber \\
    &= \sum_{j=0}^{n-1} e^{i j \Delta p_z L}  \int_0^{L} d z \, e^{i \Delta p_z z} V_w(z) \nonumber \\
    &= \mathcal{M}^{(1)} \sum_{j=0}^{n-1} e^{i j \Delta p_z L} \nonumber \\
    &= \mathcal{M}^{(1)} \frac{\sin(n \Delta p_z L/2)}{\sin(\Delta p_z L/2)} e^{i (n-1) \Delta p_z L/2} \ ,
    \label{interference}
\end{align}
where $L~(=2\pi/\kappa)$ is the spatial period of the potential, and the single wall amplitude $\mathcal{M}^{(1)}$ is given by the integral over one period of the potential.
The scattering amplitude has a characteristic interference pattern, as can be seen in \EQ{interference}.
When $\Delta p_z L = 2k \pi$, or $\Delta p_z = k \kappa$, with $k$ an integer, the superposition of the phases is a constructive mode, corresponding to the contributions from the poles at $\Delta p_z=0,\kappa$ in our analysis.
However, the other modes no longer appear, because those peaks are out of the non-adiabatic range (\ref{non-adiabatic}).
Therefore, our analysis takes account of all the dominant modes successfully. 

\section{Phenomenological implications}
\label{pheno}

Having derived the particle production rate in the presence of rippled bubble walls, we now explore its phenomenological consequences. First, we estimate the additional friction exerted on the expanding bubble wall due to the production of heavy particles and compare it with the conventional contributions. Second, we calculate the resulting relic abundance of the produced particles and demonstrate how the ripple-induced resonant channel modifies the cosmologically relevant parameter space.

\subsection{Friction on the bubble walls}

Throughout the analysis in \SEC{splitting}, we have assumed an ultra-relativistic bubble wall, $\gamma_w \gg 1$, as can be realized in sufficiently strong supercooled phase transitions. 
After bubble nucleation, the wall is accelerated by the pressure difference between the false and true vacua, denoted by $\Delta V$, while the acceleration is opposed by the friction pressure $\Delta \mathcal{P}$ exerted by the ambient plasma. 
If the driving pressure remains larger than the friction pressure, $\Delta V > \Delta \mathcal{P}$, the wall continues to accelerate until bubble collision, corresponding to the runaway regime.
Otherwise, when $\Delta V < \Delta \mathcal{P}$, the friction eventually balances the driving force and the wall reaches a terminal velocity.

Let us briefly review the plasma effect and the Next-to-Leading-Order (NLO) effect on the pressure.
In the relativistic limit, the pressure coming from the thermal plasma at the leading order is estimated as \cite{Dine:1992wr,Bodeker:2009qy}
\begin{equation}
    \Delta \mathcal{P}_{\rm LO}  \approx \sum_i \dfrac{g_i c_i \Delta m_i^2}{24}T_n^2 \ ,
\end{equation}
where $g_i$ is the number of degrees of freedom in the plasma at the nucleation temperature $T_n$, $c_i=1 \, (1/2)$ for bosons (fermions), and $\Delta m_i$ is the $i$-th component's mass shift from the symmetric phase to the broken phase.
If there is no phase-dependent vector boson involved in the phase transition, the pressure remains constant even at a high velocity. 
Thus the bubble walls can keep accelerating until collisions, and $\gamma_w$ at collision reaches the terminal one \cite{Ellis:2019oqb,Azatov:2019png,Azatov:2021ifm},
\begin{equation}
    \gamma_{w,{\rm max}} \approx \dfrac{2 R_*}{3 R_0}\lmk 1- \dfrac{\Delta \mathcal{P}_{\rm LO}}{\Delta V}\rmk \approx \dfrac{M_{\rm Pl} }{v^2 R_0} \ ,
\end{equation}
where $R_*\approx(8\pi)^{1/3}v_w /\beta(T)$ is the typical bubble size at collisions with $\beta(T)$ the inverse duration of the phase transition, and $R_0 $ is the critical bubble size at nucleation.
In the second equality, we take $\beta/H\sim \mathcal{O}(1)$.

On the other hand, 
particles charged under a gauge symmetry that is broken by the transition can emit soft gauge bosons whose mass varies across the walls, giving an additional friction known as the NLO contribution~\cite{
Bodeker:2017cim,
Gouttenoire:2021kjv,
Long:2024sqg}.
The resulting pressure is given by
\begin{equation}
    \Delta \mathcal{P}_{\rm NLO} \approx \sum_i g_i g_{\rm gauge}^3 \dfrac{vT_n^3}{16\pi^2} \gamma_w \ ,
\end{equation}
where $g_{\rm gauge}$ is the gauge coupling and $g_i$ counts the number of degrees of freedom.
Note that the $\gamma_w$ scaling behavior can stop the acceleration of the bubble wall and yield the terminal velocity.
The boost factor $\gamma_{w}$ saturates when $\Delta V \sim \Delta \mathcal{P}_{\rm NLO}$, therefore
\begin{equation}
    \gamma_{w,\rm max}\approx \min \lmk 
 \dfrac{M_{\rm Pl} }{v^2 R_0},
 \dfrac{16\pi^2}{g_i g_{\rm gauge}^3} \dfrac{\Delta V}{T_n^3 v}
    \rmk.
\end{equation}

In addition to the gauge bosons, the production of heavy particles, $\phi$, can also suppress the acceleration in our setup. 
For $n\gg 1$, the pressure is approximately estimated as
\begin{align}
    \Delta \mathcal{P}_{\phi} &= \int \frac{d^3p}{(2\pi)^3}\frac{p_z}{p^0} f_h(p)\int dP_{h\to\phi\phi}\Delta p_z \nonumber \\[1ex]
    &\approx \int \dfrac{d^3 p}{(2\pi)^3}\, \exp\left[-\dfrac{\gamma_w (E - v_w p_z)}{T_n}\right] \int \dd P_{h\to \phi\phi} \dfrac{M_\phi^2+k_\perp^2}{2x(1-x)E} \nonumber \\[1ex]
     & \approx \dfrac{V_h^2 T_n^2}{384\pi^4} \left[ e^{-\frac{M_\phi}{\gamma_wT_n}} 
     +  \dfrac{3(2n+1)}{8} e^{-\frac{M_\phi^2}{T_n \kappa \gamma_w}} +\mathcal{O}(\gamma_w^{-1}) \right]\nonumber\\
     & = \Delta \mathcal{P}_\phi^{(1)} + \Delta \mathcal{P}_\phi^{(2)} ,
\end{align}
where we assume the thermal equilibrium distribution of the Higgs, $f_h(p)$, as the Boltzmann distribution. 
In the second equality, for the $|\mathcal{M}|_1^2$ contribution, we have approximated the average momentum transfer by $M_\phi^2/(2E)$ which is independent of the final-state energy momentum since $\Delta p_z\approx 0$.
For the $|\mathcal{M}|_2^2$ contribution, we have substituted $\Delta p_z = \kappa$ by definition.
The first term originates from $|\mathcal{M}|^2_1$ and is consistent with the result of the previous work \cite{Azatov:2020ufh}, and the second term is the contribution from $|\mathcal{M}|^2_2$.

The non-adiabatic condition (\ref{non-adiabatic}) is satisfied in the high $\gamma_w$ range, $\gamma_w\gg M_\phi^2/\kappa T_n\gg M_\phi/T_n$.
In this regime, the two contributions take the values $\Delta \mathcal{P}_\phi^{(1)} \approx V_h^2T_n^2/(384\pi^2)$ and $\Delta \mathcal{P}_\phi^{(2)} \approx n V_h^2T_n^2/(512\pi^2)$, respectively, where we assume $n\gg 1$.
We can compare the new contribution with the LO and NLO effects, whose ratios are given by
\begin{align}
\frac{\Delta \mathcal{P}_{\rm LO}}{
\Delta \mathcal{P}_\phi^{(2)}
} \simeq \sum_i 8\pi^4 \frac{\Delta m_i^2}{nV_h^2} \ , \qquad 
\frac{\Delta \mathcal{P}_{\rm NLO}}{
\Delta \mathcal{P}_\phi^{(2)}
}\simeq \sum_i 12\pi^2 g_{\rm gauge}^3 \frac{T_n}{n\lambda V_h}\gamma_w \, .
\end{align}
In our setup, there exists a contribution from $\lambda \phi^2 |H|^2$ to the LO friction.
Taking this into account, the first equality reads $\Delta \mathcal{P}_{\rm LO}/\Delta \mathcal{P}_\phi^{(2)} \gtrsim 8\pi^4 \lambda v^2 / n(\lambda v)^2 \simeq 8\pi^4/n\lambda$.
As we will discuss in \SEC{sec:evolution}, $n$ is typically at most $100$, and therefore the new contribution never dominates the LO friction as long as we take $\lambda$ within a perturbative range.
Regarding the second equality as well, considering that $\gamma_w$ takes huge values in supercooled phase transitions in general, the new contribution does not dominate the NLO friction unless the gauge coupling is extremely small.

\subsection{Relic abundance}

Let us now estimate the relic abundance of $\phi$ produced via the splitting process, $h\to\phi\phi$.
Note that the following estimate neglects subsequent shell evolution and applies only where free streaming remains a valid approximation~\cite{Baldes:2024wuz}.
In the plasma frame, the number density of $\phi$ is
\begin{equation}
    n_{\phi} \approx \dfrac{2}{\gamma_w v_w} \int \dfrac{d^3 p}{(2\pi)^3} P_{h\to \phi\phi} \exp\left[-\dfrac{\gamma_w (E - v_w p_z)}{T_n}\right],
\end{equation}
where we have assumed the Boltzmann distribution for $h$ as with the previous subsection.
For $n\gg1$, the contribution from $P_{h\to \phi\phi}^{\Delta p_z \approx 0}$ takes the form,
\begin{align}
  n_\phi^{\Delta p_z\approx 0} =& \, \dfrac{2}{\gamma_w v_w} \int \dfrac{d^3 p}{(2\pi)^3} \dfrac{V_h^2}{384\pi^2 M_\phi^2} 
    \Theta(E-2M_\phi) e^{-\frac{\gamma_w (E - v_w p_z)}{T_n}} \nonumber \\[1ex]
    =& \, \dfrac{V_h^2}{768\pi^4 M_\phi^2 v_w \gamma_w}\int_{2M_\phi}^\infty d E \, E e^{-\frac{\gamma_w E}{T_n}} \int_{-E}^E d p_z \, e^{\frac{\gamma_w v_w p_z}{T_n}}\nonumber \\[1ex]
    =& \, \dfrac{V_h^2 T_n^3 e^{-\frac{2M_\phi\gamma_w (1-v_w)}{T_n}}}{768\pi^4 M_\phi^2 v_w^2 \gamma_w^3} \left[ 
    \dfrac{2M_\phi}{(1-v_w)T_n} +\dfrac{1}{(1-v_w)^2 \gamma_w}
    \right] \nonumber \\[1ex]
     \approx& \, \dfrac{V_h^2 T_n^3}{192\pi^4 M_\phi^2} e^{-\frac{M_\phi}{T_n\gamma_w}} \left[
    1+\dfrac{M_\phi}{T_n\gamma_w}+\mathcal{O}(\gamma_w^{-2}) \right].
\end{align}
Here we have used $\gamma_w v_w = \sqrt{\gamma_w^2-1}\approx\gamma_w - 1/(2\gamma_w)$ for $\gamma_w\gg1$.
This is consistent with the previous result \cite{Azatov:2021ifm} up to the prefactor.
The contribution from $P_{h\to \phi\phi}^{\Delta p_z \approx \kappa}$ has the form,
\begin{align}
    n_\phi^{\Delta p_z\approx \kappa} =& \, \dfrac{2}{\gamma_w v_w} \int \dfrac{d^3 p}{(2\pi)^3}  \dfrac{(2n+1)V_h^2}{512 E \kappa} 
       \sqrt{1-\dfrac{2M_\phi^2}{E \kappa}} \Theta\lmk E-\dfrac{2M_\phi^2}{\kappa}\rmk \Theta(E-2M_\phi) e^{-\frac{\gamma_w (E - v_w p_z)}{T_n}} \nonumber \\[1ex]
    \approx& \, \dfrac{(2n+1)V_h^2}{2048\pi^4 M_\phi^2 v_w \gamma_w}\int_{2M_\phi}^\infty d E \, E \lmk \dfrac{2M_\phi^2}{E \kappa}\rmk e^{-\frac{\gamma_w E}{T_n}} \int_0^E d p_z \, e^{\frac{\gamma_w v_w p_z}{T_n}} \nonumber \\[1ex]   
    \approx& \, \dfrac{(2n+1)V_h^2 T_n^2}{512\pi^4 \kappa \gamma_w  } e^{-\frac{M_\phi^2}{T_n \kappa \gamma_w}}
    +\mathcal{O}(\gamma_w^{-2}) \ ,
\end{align}
where we have used $\sqrt{1-a/x} \, \Theta(x-a)\approx\Theta(x-a)$.
As expected, the abundance is adiabatically suppressed in the range of $\gamma_w\lesssim M_\phi^2/\kappa T_n$ by $e^{-M_\phi^2/\kappa\gamma_wT_n}$. We note that the ripple effect gets suppressed by $\gamma_w^{-1}$, because it is out of the resonance region, $\Delta p_z > \kappa$, for a large incident energy $E\simeq \gamma_w T_n$.

The total number density of non-thermal accumulation at $T=T_n$ is given by
\begin{align}
    n_\phi &= n_\phi^{\Delta p_z\approx 0} + n_\phi^{\Delta p_z\approx \kappa}\nonumber\\[1ex]
    &\approx \dfrac{V_h^2 M_\phi}{192\pi^4} \left[ e^{-\frac{M_\phi}{T_n\gamma_w}}\dfrac{T_n^3}{M_\phi^3} + 
    \lmk \dfrac{T_n^2}{M_\phi^2} e^{-\frac{M_\phi}{T_n\gamma_w}}+\dfrac{3(2n+1)T_n^2}{8\kappa M_\phi} e^{-\frac{M_\phi^2}{T_n\kappa \gamma_w}}\rmk  \gamma_w^{-1} +\mathcal{O}(\gamma_w^{-2}) \right].
\end{align}
One can see from this result that the ripple effect can enhance the abundance significantly, if the number of ripples is large enough, $n\gtrsim \kappa M_\phi/T_n^2$, with the non-adiabatic condition satisfied.
Assuming no subsequent thermalization, annihilation and extra dilution, the relic abundance reads
\begin{align}
    \Omega_{\phi} h^2 &= \dfrac{M_\phi n_\phi}{\rho_c / h^2} \dfrac{g_\star(T_0) T_0^3}{g_\star (T_{\rm reh}) T_{\rm reh}^3}\nonumber\\[1ex]
    &\approx 3\times10^4 \frac{1}{g_\star(T_{\rm reh})} \lmk \dfrac{\lambda^2 v}{M_\phi} \rmk \lmk\frac{v}{\rm GeV}\rmk \lmk\frac{T_n}{T_{\rm reh}}\rmk^3 \lmk 1 + \dfrac{3(2n+1)M_\phi^2}{8 \kappa \gamma_w T_n}\rmk,
    \label{eq:relic-abundance}
\end{align}
where $\rho_c$ denotes the critical energy density, $h$ is the reduced Hubble constant, $T_0$ is the temperature today, $T_{\rm reh}$ is the reheating temperature, and $g_\star(\cdot)$ is the entropy number of degrees of freedom at the given temperature.
Here, we have used $e^{-M_\phi/T_n \gamma_w}\approx 1-M_\phi/T_n \gamma_w$ and $e^{-M_\phi^2/T_n \kappa \gamma_w}\approx 1{-M_\phi^2/T_n \kappa \gamma_w}$, and ignored the $\mathcal{O}(\gamma_w^{-1})$ contribution from $|\mathcal{M}|^2_1$.
The abundance is significantly suppressed by $(T_n/T_{\rm reh})^3$ for a strongly supercooled phase transition.

\begin{figure}
    \centering
    \includegraphics[width=0.9\linewidth]{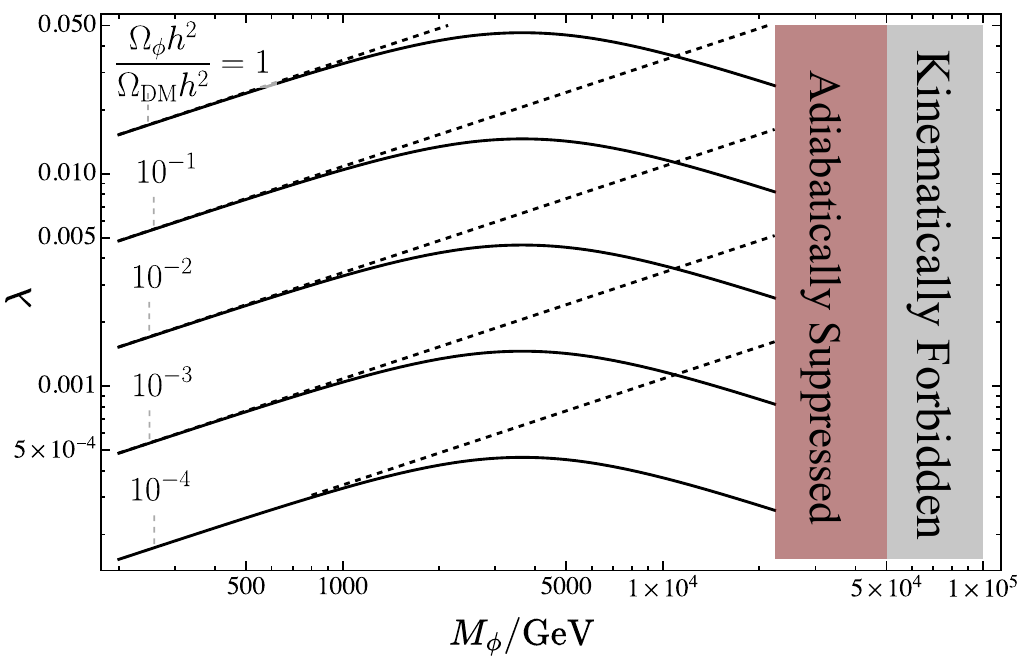}
    \caption{The relic abundance of $\phi$ produced by bubble expansion.
    The black dashed lines represent the abundance with ripple number $n=0$,
    while the black solid lines represent the abundance with ripple number $n=100$. 
    Parameters are set to be $v=T_n=200 \, \rm GeV$, $g_*=100$, $T_{\rm reh}/T_n=10$, $\gamma_w=500$ and $\kappa=10 \, \rm TeV$. The red-shaded region does not satisfy the non-adiabatic threshold (\ref{non-adiabatic}), while the production is not kinematically allowed in the gray-shaded region.
    Each result is estimated for $\Omega_\phi/\Omega_{\rm DM} = 1, 10^{-1}, 10^{-2}, 10^{-3}, 10^{-4}$ with $\Omega_{\rm DM}h^2\simeq 0.12$ the observed DM abundance.
    }
    \label{fig:relic-abundance}
\end{figure}

\FIG{fig:relic-abundance} shows the relic abundance in the plane of $(M_\phi, \lambda)$.
Here we take $v=T_n=200 \, \rm GeV$,  $T_{\rm reh}/T_n=10$, $g_*(T_{\rm reh})=100$, $\gamma_w=500$, and $\kappa=10 \, \rm TeV$.
The black solid and dashed lines represent the abundance with ripple number $n=100, 0$, respectively. 
The red-shaded region does not satisfy the non-adiabatic threshold (\ref{non-adiabatic}), while the production is not kinematically allowed in the gray-shaded region.
Each result is estimated for $\Omega_\phi/\Omega_{\rm DM} = 1, 10^{-1}, 10^{-2}, 10^{-3}, 10^{-4}$ with $\Omega_{\rm DM}h^2\simeq 0.12$ the observed DM abundance.
One can see from the figure that the required $\lambda$ is suppressed around the non-adiabatic threshold, because this range corresponds to the peak $\Delta p_z\approx\kappa$.

\section{Evolution of the Higgs field
\label{sec:evolution}}

At the bubble nucleation, the tunneling process of $h$ terminates at the exit point $h_0$, which is much smaller than the value at the true vacuum in a strongly supercooled phase transition.
Due to the near flatness of the potential at $h\approx h_0$, the Higgs field slowly starts to move from the exit point and finally oscillates around the true vacuum.
In this section, we estimate the number of the Higgs oscillations, or equivalently, that of the ripples on the bubble wall, $n$, which is the enhancement factor for the particle splitting.

Although the finite-temperature bounce is only $O(3)$ symmetric, the post-nucleation evolution of a strongly supercooled bubble at the initial stage is expected to be dominated by the vacuum pressure.
The ripples relevant to our analysis are generated during this initial expansion stage.
Even if plasma friction later drives the wall toward a terminal velocity, this is expected to occur only after the initial accumulation of the ripples; the plasma then can affect how strongly the accumulated ripples get compressed and how rapidly they get damped.
Since our purpose here is to estimate the number of ripples generated during the initial expansion, we approximate the expanding bubble as $SO(1, 3)$ symmetric.
The subsequent plasma effects and the limitations of this treatment are discussed in Sec.~\ref{conclusion}.

\subsection{Effective number of Higgs oscillations}

To estimate the effective number of the ripples, let us consider the effective potential of $h$ separately in two stages: the slow-rolling stage (denoted by $h_1$) and the oscillation stage (denoted by $h_2$).
For the slow-rolling stage, the effective potential can be parameterized by (almost) conformally flat shape with a dimensionless coupling $\lambda_h$,
\begin{equation}
    V^{(1)}_{\rm eff} = -\dfrac{\lambda_h}{4} h_1^4 + \text{const}.
    \label{slow-rolling}
\end{equation}
In the oscillation stage, $h$ feels a quadratic potential around the true vacuum at $\langle h \rangle = v$,
\begin{equation}
    V^{(2)}_{\rm eff} = \frac{1}{2}m^2 (h_2-v)^2 \ ,
\end{equation}
with an effective mass $m$.
Under the approximations described above, the evolution of $h_i$ is governed by the following $SO(1,3)$-symmetric equation of motion
\begin{equation}
    \dfrac{d^2}{d \tau^2} h_{i} + \dfrac{3}{\tau}\dfrac{d}{d \tau} h_{i} + \dfrac{d V^{(i)}_{\rm eff}}{d h_i} =0 \ ,
    \label{evolution_eq}
\end{equation}
where $i=1,2$, and we define the intrinsic time $\tau = \sqrt{t^2-r^2}$.

For the slow-rolling stage, \EQ{evolution_eq} is equivalent to the equation giving the Lorentzianized Fubini instanton solution \cite{Fubini:1976jm,Lipatov:1976ny}. Thus we obtain the exact analytic solution,
\begin{equation}
    h_1(\tau) = \dfrac{h_0}{1-\lambda_h h_0^2 \tau^2 /8} \ .
\end{equation}
Here we consider the exit point $h_0$ as a free parameter, which should be fixed by the breaking of scale invariance.
In the next subsection, we consider a concrete model to estimate the number of ripples numerically by determining $h_0$. 
Since the approximation (\ref{slow-rolling}) works only in the range of $h_1\ll v$, $h_1(\tau_0) \sim v$ can give a naive estimate of the endpoint in the slow-rolling stage as
\begin{equation}
    \tau_0 \approx \frac{1}{h_0} \sqrt{\dfrac{8}{\lambda_h} \lmk 1- \dfrac{h_0}{v}\rmk} \ .
\end{equation}
At $\tau>\tau_0$, $h$ goes into the oscillation stage. 
The equation of motion with the quadratic potential has an exact solution,
\begin{equation}
    h_2(\tau) = v-\frac{1}{\tau} \lmk  A J_1(m \tau) + B Y_1(m \tau) \rmk,
\end{equation}
where $A, B$ denote constants, and $J_1, Y_1$ are the Bessel functions of the first and second kind, respectively.
The constants $A, B$ can be determined by the boundary conditions, $h_1(\tau_0) = h_2(\tau_0)$ and $h_1'(\tau_0) = h_2'(\tau_0)$.
In the asymptotic regime $m\tau\gg 1$, the Bessel functions are given by
\begin{align}
    J_1(m\tau) \approx \sqrt{\dfrac{2}{\pi m \tau}} \cos\lmk m \tau - \dfrac{3\pi}{4} \rmk,\\[1ex]
    Y_1(m \tau) \approx \sqrt{\dfrac{2}{\pi m \tau}} \sin\lmk m \tau - \dfrac{3\pi}{4} \rmk,
\end{align}
which shows that the amplitude of $h_2(\tau)$ decays as $\tau^{-3/2}$ with the oscillation frequency $m$.

We approximately count the transient number of the ripples at $\tau$ by $n(\tau)\approx m(\tau-\tau_0)/(2\pi)$.
By changing $\tau_0$ via parameters $\lambda_h$ and $h_0$, the effective number of ripples $n$ is estimated as
\begin{equation}
    n_{\rm eff} \equiv \dfrac{\int_{\tau_0}^\infty d \tau \ n(\tau) (h_2(\tau)-v)^2}{\int_{\tau_0}^\infty d \tau  (h_2(\tau)-v)^2} 
    \approx \dfrac{m \tau_0}{2\pi}
    \approx \dfrac{m}{2\pi h_0}  \sqrt{\dfrac{8}{\lambda_h} \lmk 1- \dfrac{h_0}{v}\rmk} \ .
    \label{n_eff}
\end{equation}
As the Higgs stays at $h_0$ for a longer time for a stronger supercooled phase transition, the decay of the oscillation amplitude is suppressed, resulting in a larger number of ripples.

\subsection{A model: classical scale invariance}

A supercooled phase transition can be realized in the presence of approximate scale invariance. Small breaking of scale invariance has been implemented in various models, such as conformal field theory \cite{Creminelli:2001th,vonHarling:2017yew,Baratella:2018pxi,Fujikura:2019oyi,Fujikura:2025iam,Agrawal:2025wvf} and classical scale invariance \cite{Witten:1980ez,Iso:2009ss,Iso:2009nw,Konstandin:2011dr,Jinno:2016knw,Jaeckel:2016jlh,Iso:2017uuu,Marzo:2018nov,Hambye:2018qjv,Ellis:2019oqb,Ellis:2020nnr,Dasgupta:2022isg,Sagunski:2023ynd}. Here we take a model with classical scale invariance as in \cite{Rescigno:2025ong} to estimate $n_{\rm eff}$ more explicitly.

We consider scale invariance which is broken only by radiative symmetry breaking (RSB),
so that the nucleation rate changes on temperature solely via the slowly running beta function of the flat direction,
which significantly prolongs the false vacuum trapping time.
In this case, any scale should be generated only through loop-induced effects rather than at the tree-level.
Then,  the most general Lagrangian is \cite{Rescigno:2025ong}
\begin{equation}
    \mathcal{L} = -\frac{1}{4} F_{\mu\nu}^a F^{\mu\nu}_a + \dfrac{1}{2} D\varphi^i D\varphi_i + i \bar\psi_s \cancel{D} \psi^s -\frac{1}{2} (Y_{i s'}^s  \bar\psi_s \varphi^i\psi^{s'} + {\rm h.c.}) - \frac{\lambda_\varphi^{ijkl}}{4}\varphi_i \varphi_j \varphi_k \varphi_l \ ,
\end{equation}
where $n_{\rm S}$ real scalars $\varphi$, $n_{\rm F}$ Weyl fermions $\psi$, and $n_{\rm A}$ vectors $A$ (with the field strength $F$, the covariant derivative $D_\mu = \partial_\mu -i g_{\rm gauge} \tau_a A^a_\mu$,  gauge coupling $g_{\rm gauge}$ and the symmetry group generator $\tau_a$) are considered.
Note that the Yukawa matrix $Y_{s'}^s$ is symmetric and the quartic coupling $\lambda_\varphi$ is totally symmetric.

In the RSB theory, a real scalar develops a flat direction parameterized as
\begin{equation}
    \varphi_{\rm flat}^i =h\, \hat \varphi_{\rm flat}^i \ .
\end{equation}
Here $\hat \varphi_{\rm flat}$ is a unit vector in the scalar field space satisfying $\hat \varphi_{\rm flat}^i \hat \varphi_{{\rm flat},i}=1$, and $h$ is the value of the classical background in the direction $\hat \varphi_{\rm flat}$.
The renormalization group (RG)-improved potential for $h$  at the renormalization scale $\mu$ is
\begin{equation}
    V_q(h) = \dfrac{\lambda_h(\mu)}{4}h^4 \ ,
\end{equation}
where the self-coupling $\lambda_h$ is defined as
\begin{equation}
    \lambda_h(\mu) = \lambda_\varphi^{ijkl}(\mu) \hat \varphi_{\rm flat}^i \hat \varphi_{\rm flat}^j \hat \varphi_{\rm flat}^k \hat \varphi_{\rm flat}^l \ .
\end{equation}
The conditions of $h$ that corresponds to a flat direction on the renormalization scale $\mu_0$, and that $V(h)$ has an extreme are
\begin{align}
    &\lambda_h(\mu_0) = \lambda_\varphi^{ijkl}(\mu_0) \hat \varphi_{\rm flat}^i \hat \varphi_{\rm flat}^j \hat \varphi_{\rm flat}^k \hat \varphi_{\rm flat}^l =0 \ , \\[1ex]
 &\lambda_\varphi^{ijkl}(\mu_0)  \hat \varphi_{\rm flat}^j \hat \varphi_{\rm flat}^k \hat \varphi_{\rm flat}^l =0 \ ,\label{eq:Vq-extreme}
\end{align}
respectively.
The one-loop quantum effective potential renormalized at $\mu_0$ becomes
\begin{equation}
   V_q(h) = \dfrac{\beta_{\lambda_h}}{4}h^4 \lmk \log \dfrac{h}{\vev h}-\frac14 \rmk,
\end{equation}
where 
\begin{equation}
    \beta_{\lambda_h} =\left. \mu \dfrac{\dd \lambda_h}{\dd \mu} \right|_{\mu=\mu_0} 
\end{equation}
represents the one-loop beta function of $\lambda_h$, and $\vev h$ is the VEV of $h$ introduced by dimensional transmutation which is set by $\mu_0$ in a renormalization scheme dependent way.
Nevertheless, the way of choosing renormalization scheme does not affect $\vev h$.
We also note that $\beta_{\lambda_h}>0$ ensures that $\vev h$ is a minimum of $V_q$.

The beta function of $\lambda_h$ runs as
\begin{align}
    (4\pi)^2 \beta_{\lambda_h} =& \, 3! \hat \varphi_{\rm flat}^i \hat \varphi_{\rm flat}^j \hat \varphi_{\rm flat}^k \hat \varphi_{\rm flat}^l \nonumber \\
& \times \lmk 
 N_{\rm S} \lambda^{mni(j}_\varphi \lambda^{kl)mn}_\varphi
 +\frac{N_{\rm V}}4 g_{\rm gauge}^4 \{\tau^a\tau^b\}_{i(j}\{\tau^a\tau^b\}_{kl) } +N_{\rm F}\tr(Y_i Y^\dagger_{(j} Y_k Y^\dagger_{l)} ) 
\rmk \nonumber \\
=&\, g^4-2y^4 \ ,
\end{align}
where $N_{\rm S}=1, N_{\rm V}=3$, and $N_{\rm F}=-2$ take account of the degrees of freedom of each real scalar, vector boson, and Weyl fermion (with opposite sign for spin $1/2$ statistics),  respectively.
The notation $(jkl)$ denotes the normalized total symmetric permutation for indices $j,k$, and $l$; 
$\{AB\} = AB+BA$ is the anti-commutator,
and in the last line we have introduced the effective portal couplings $g$ and $y$ for bosons and fermions, respectively.
Note that the running of anomalous dimension of $\varphi$ proportional to $\lambda_\varphi^{ijkl}(\mu_0)  \hat \varphi_{\rm flat}^j \hat \varphi_{\rm flat}^k \hat \varphi_{\rm flat}^l$ becomes irrelevant due to \EQ{eq:Vq-extreme}.

The tree-level potential $V_0$ can be expressed as
\begin{equation}
    V_0 = \frac12 \varphi_{\perp}^i M^2_{{\rm S},ij}(h)  \varphi_{\perp}^j 
    + \frac{1}{2} A^a_\mu M^2_{{\rm V},ab}(h) A^{b,\mu} 
    + \bar\psi_s M_{{\rm F},s'}^s (h) \psi^{s'} + \mathcal{O}(\varphi_\perp^2, A^2, \bar\psi\psi) \ ,
\end{equation}
where $\varphi^i_\perp \varphi_{{\rm flat},i}=0$, and
\begin{align}
     M^2_{{\rm S},ij}(h)  &= \frac12 \lambda_{{\varphi, ijkl}}  \hat \varphi_{\rm flat}^k  \hat \varphi_{\rm flat}^l h^2 \ ,\\[1ex]
     M^2_{{\rm V},ab}(h)  &=g_{\rm gauge}^2 \tr\lmk  \hat \varphi_{\rm flat}^T\tau_a \tau_b \hat \varphi_{\rm flat}\rmk h^2 \ ,\\[1.5ex]
     M_{{\rm F},s'}^s (h) &= Y_{i s'}^s \hat \varphi^i h \ .
\end{align}
Here, we note that
$n_{\rm S} = \dim M_{\rm S}$, 
$n_{\rm A} = {\rm rank} \, M_{\rm V}$, and
$n_{\rm F} = \dim M_{\rm F}$.
Then the thermal correction in the Matsubara formalism is defined as
\begin{align}
    V_\beta (h) = & \, \dfrac{T^4}{2\pi^2}
    \sum_{i={\rm S, V, F}}
    N_i \tr J_i
    \lmk  
    \dfrac{|M_i|^2}{T^2}
    \rmk  \nonumber\\
  & - \dfrac{T}{12\pi}\sum_{j={\rm S, V}}N_j \tr\lmk
    (|M_j|^2+\Pi_j)^{3/2}-|M_j|^3
    \rmk  \nonumber  \\
   \simeq& -\lmk n_{\rm B}-\frac78 n_{\rm F} \rmk\dfrac{\pi^2 T^4}{90}+ \dfrac{(g^2-2y^2) T^2}{24} h^2 - \dfrac{T}{12\pi}\lmk g^2 h^2 + \epsilon^2 T^2 \rmk^{3/2} \nonumber \\[1ex]
    &- \dfrac{g^4-2y^4}{64\pi^2} h^4  \lmk\log\dfrac{T^2}{(g^2-2y^2)h^2} -C\rmk,
\end{align}
where $T$ is the temperature introduced from the Matsubara formalism of finite temperature field theory, and in the second equality, we take the high-$T$ approximation.
The thermal bosonic and fermionic functions are defined as
\begin{align}
    J_{\rm S}(x) = J_{\rm V}(x) =\int^\infty \dd y\, y^2\log \lmk 1- e^{-\sqrt{y^2+x}} \rmk,\\
    J_{\rm F}(x) = \int^\infty \dd y\, y^2\log \lmk 1+ e^{-\sqrt{y^2+x}} \rmk ,
\end{align}
and $\Pi_j$'s are the leading part of the $T$ dependent self-energies that can be parameterized as  $\epsilon^2 T^2$.
Note that the dependence of $\Pi_j$ originates from the daisy resummation accounting for higher-loop thermal corrections around the critical temperature.
$C=3/2$ is the constant that coincides with the renormalization scheme dependent constant in the $\overline{\rm DR}$ scheme for both bosons and fermions.
We also note that the $T^4$ term is the Stefan-Boltzmann free-energy of $n_{\rm B}=n_{\rm S}+3n_{\rm A}$ bosonic and $n_{\rm F}$ fermionic degrees of freedom.

\begin{figure}
    \centering
    \includegraphics[width=0.8\linewidth]{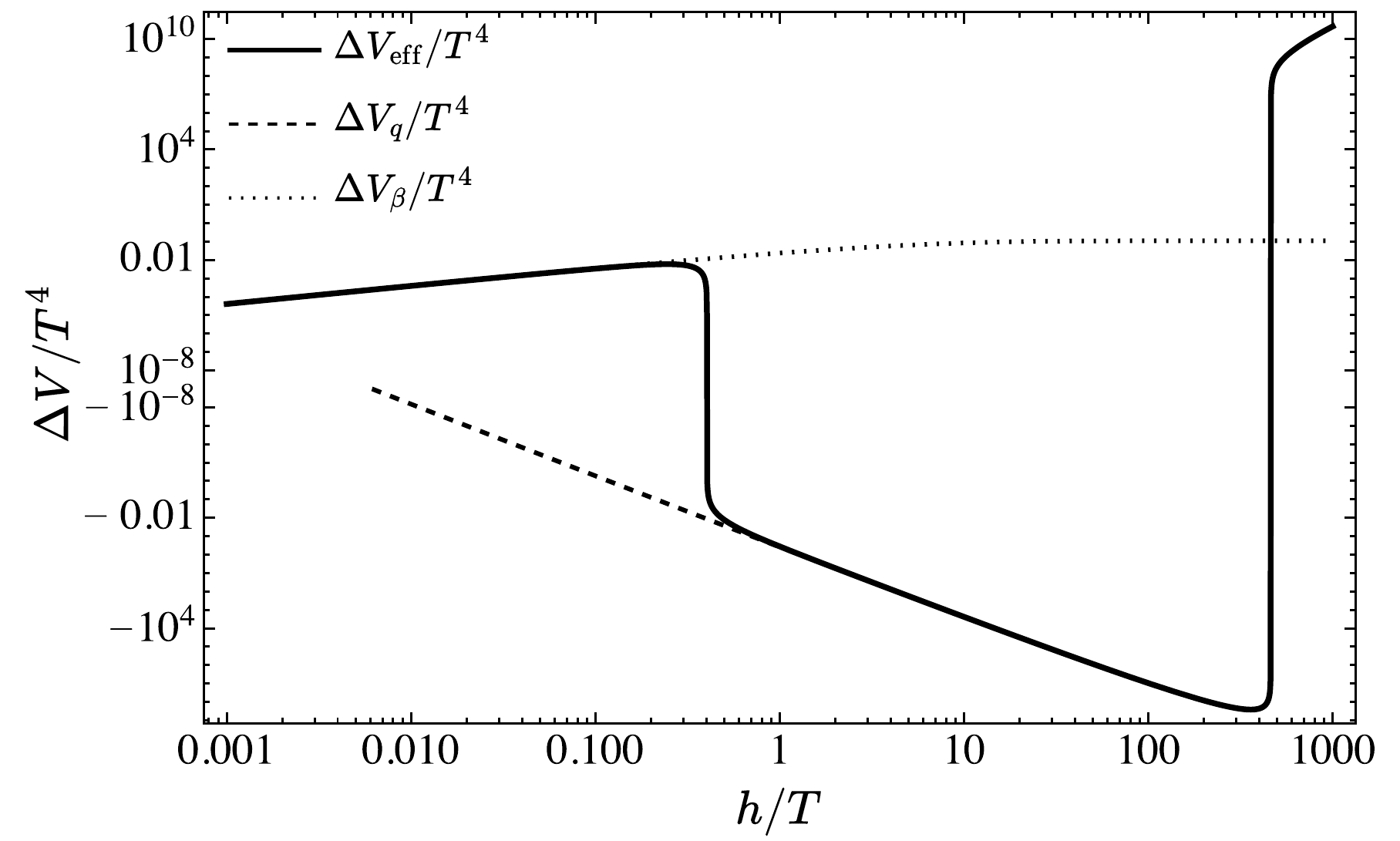}
    \caption{Relative potential energy $\Delta V(h) =  V(h) -  V(0)$ with $\epsilon=y=0$ and $g=1$. 
    The black solid line corresponds to the exact effective potential, while the dotted (dashed) line corresponds to $V_\beta$ ($V_{q}$).
    }
    \label{fig:veff}
\end{figure}

After including both quantum correction $V_{q}$ and thermal correction $V_{\beta}$, the effective potential of the flat direction field $h$ takes the form,
\begin{equation}
    V_{\rm eff}(h) = V_q(h) + V_{\beta}(h) \ .
\end{equation}
While the form of $V_{\rm eff}$ can be numerically obtained, it is helpful to introduce its high-temperature and low-temperature approximation.
In the low-temperature limit, thermal corrections are exponentially suppressed and the Coleman-Weinberg radiative correction takes the dominance,
\begin{equation}
    V_{\rm eff}(h\gg T) \approx V_q (h)= \dfrac{g^4-2y^4}{64\pi^2} h^4 \lmk \log \dfrac{h}{ \langle h\rangle}-\frac14 \rmk.\label{eq:veff-low}
\end{equation}
In the high-temperature limit, on the other hand, the effective potential of $h$ takes the form,
\begin{align}
    V_{\rm eff}(h\ll T) \approx& -\lmk n_{\rm B}-\frac78 n_{\rm F} \rmk\dfrac{\pi^2 T^4}{90}+ \dfrac{(g^2 -2y^2)T^2 }{24} h^2 \nonumber\\[1ex]
    &- \dfrac{T}{12\pi}\lmk g^2 h^2 + \epsilon^2 T^2 \rmk^{3/2} + \dfrac{g^4-2y^4}{64\pi^2} h^4  \log\dfrac{T^2}{\mu_0^2}\label{eq:veff-high} \ .
\end{align}
\FIG{fig:veff} shows the shape of $V_{\rm eff}$.
The black solid line corresponds to the exact effective potential, while the dotted (dashed) line corresponds to $V_\beta$ ($V_{q}$).
Typically, the thermal correction $V_{\beta}$  controls the size of the potential barrier in the regime of $h\ll T$, 
while the quantum correction $V_{q}$ is responsible for the formation of the true vacuum, where the VEV $\langle h \rangle$ is expected to be much larger than the temperature for a supercooled phase transition.
These two regimes are therefore described by the high and low temperature limits, respectively.

\begin{figure}
\hspace{-1cm}
    \centering
    \includegraphics[width=0.85\linewidth]{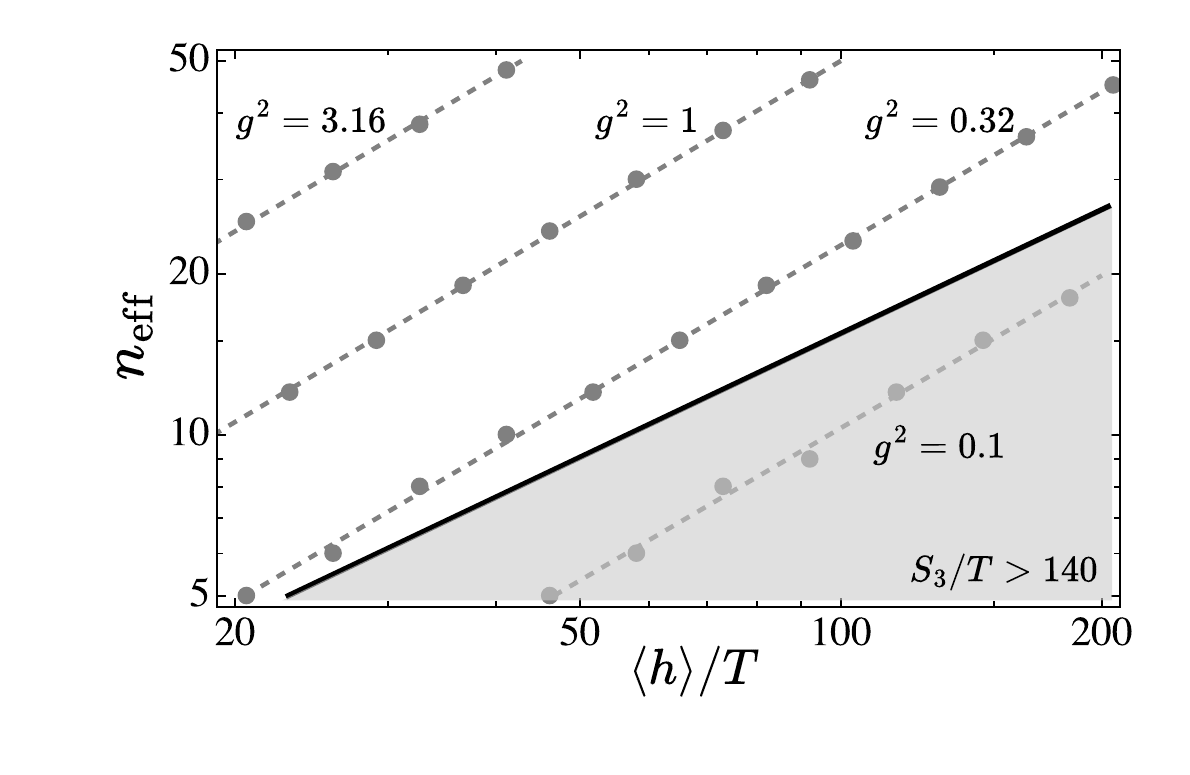}
    \vspace{-0.5cm}
    \caption{Effective ripple number $n_{\rm eff}$ with parameters set to be $\epsilon=y=0$. 
    The gray dashed lines and dots correspond to the cases for $g^2=0.1$ to $g^2=3.16$ with different temperature. The black line shows the effective ripple number at the nucleation temperature where the bounce action reaches $S_3/T=140$, while the gray shaded region is excluded by inefficient nucleation during a Hubble time inside a Hubble patch. 
    Below the nucleation temperature, the nucleation continues until the percolation.
    }
    \label{fig:neff}
\end{figure}

The evolution of $h$ is described as discussed in the previous subsection by taking the bounce solution as the initial condition at the nucleation time $t=t_n$, which is identified with  the exit point $h_0$.
In the finite temperature field theory, the bounce equation is obtained by minimizing the dimensionless Euclidean action $S_3/T$ of classical profile $h(\rho)$ where $\rho$ is the spatial radius from the center of the bubble,
\begin{equation}
    \partial_\rho^2 h + \dfrac{2}{\rho}\partial_\rho h  - \frac{dV_{\rm eff}}{dh}= 0 \ .
\end{equation}
By numerically solving this equation with the effective potential (without using the high and low-temperature approximations), we count the effective ripple number $n_
{\rm eff}$ as shown in \FIG{fig:neff}.
The gray dashed lines and dots correspond to the effective ripple number for $g^2=0.1 \sim 3.16$. 
The black line shows the results at the nucleation temperature $T=T_n$,
where the bounce action reaches $S_3/T=140$ which corresponds to the phase transition at the $\mathcal{O}(1) \, {\rm TeV}$ scale. 
The shaded region corresponds to inefficient nucleation during the Hubble time inside a Hubble patch.
The increase with $\langle h\rangle/T$ is consistent with the parametric dependence of \EQ{n_eff}, since $m/h_0 \sim \langle h \rangle/T$.
We also note that decreasing the coupling flattens the potential near the exit point, resulting in a larger number of ripples.
The numerical estimate indicates that $n_{\rm eff}= \mathcal{O}(10-100)$ is a typical value for significant supercooling.

\section{Discussion and conclusions}
\label{conclusion}

In the present study, we have explored non-thermal particle production induced by ultra-relativistic bubble walls with an oscillatory internal structure. Extending the bubble-expansion mechanism proposed in previous studies, we modeled the bubble wall by a thick-wall profile containing multiple ripples and investigated how these ripples modify the production of heavy particles coupled to the Higgs field. Our analysis shows that the ripple structure introduces a new resonant momentum-transfer channel in addition to the conventional contribution associated with smooth bubble walls. The repeated oscillations of the wall profile lead to constructive interference in the scattering amplitude, enhancing the production probability when the momentum transfer matches the characteristic ripple scale. Consequently, the abundance of non-thermally produced particles can be significantly enhanced compared with the smooth-wall case.  To see that the ripple structures are expected to arise in realistic phase transitions, we investigated the post-tunneling evolution of the Higgs field. Since the tunneling endpoint in strongly supercooled phase transitions is typically far from the true-vacuum minimum, the field naturally overshoots the minimum and undergoes coherent oscillations. These oscillations leave a sequence of ripples on the expanding bubble wall. We have developed both analytical and numerical frameworks to estimate the effective number of ripples in a classically scale-invariant model, demonstrating that a non-negligible number of ripples can naturally be generated. Overall, it is concluded that the internal structure of bubble walls can play an important role in non-thermal particle production.

In this work, we have modeled the bubble-wall profile by a simple oscillatory ansatz to obtain the analytical insight into non-thermal particle production. Although the specific functional form is idealized, we can expect the qualitative conclusion to be more general. The essential ingredient responsible for the resonant enhancement is the presence of oscillatory Fourier components in the bubble-wall profile, rather than the precise shape of the oscillations. Hence, any realistic bubble wall containing coherent oscillatory structures is expected to exhibit similar resonant momentum-transfer channels. Nevertheless, quantitative predictions, including the resonance width, the enhancement factor, and the overall production rate, generally depend on the detailed Fourier spectrum of the wall profile. A more complete treatment would therefore apply the particle-production formalism developed here to more realistic wall profiles obtained in concrete models, allowing the robustness and magnitude of the resonant enhancement to be assessed beyond the specific analytic ansatz.

An important issue beyond the present analysis is the stability of the ripple structure.
We have not taken account of the dynamical evolution of the ripple structure.
In reality, however, the coherent oscillations can be damped through interactions with the ambient plasma and the backreaction effect associated with heavy-particle production.
It is not a trivial issue whether the ripples survive during the phase transition, while the resonant enhancement only requires that the oscillatory structure remain coherent during some period in which incident particles traverse multiple ripples.
A quantitative assessment of the subsequent shell evolution and its impact on the wall friction and relic abundance~\cite{Baldes:2024wuz} is left for future work.

The role of the plasma has two competing effects. On the one hand, a supercooled transition reduces the plasma density and may weaken the damping of the scalar oscillations. On the other hand, the same plasma provides the incident particles responsible for heavy-particle production. Consequently, stronger supercooling does not necessarily imply more efficient particle production. Moreover, the ripple-induced channel is expected to be the most effective when the incident energy becomes comparable to the resonance scale, $\gamma_{w} T_{n} \sim {M_\phi^2}/{\kappa}$. At earlier times the incident energy is insufficient to efficiently excite the resonant channel, whereas at much larger Lorentz factors the system moves away from the resonance.

We expect that the present analysis provides a reasonable leading-order description in the parameter region where the energy transferred to produced particles remains small compared with that stored in the ripple configuration. Indeed, our estimate indicates that heavy-particle production typically provides only a subdominant contribution to the total friction acting on the bubble wall. Nevertheless, this result does not necessarily guarantee the stability of the oscillatory component. A quantitative assessment would determine the lifetime and coherence of the ripple structure, which is left for a future study.

The present work has focused on particle production from expanding ultra-relativistic bubble walls before bubble collisions. However, bubble collisions can also provide an additional source of non-thermal particle production. A quantitative comparison between particle production during the bubble-expansion and bubble-collision stages would provide a more complete understanding of non-thermal particle production in first-order phase transitions and is left for a future work.

\section*{Acknowledgments}

R.J. is grateful to Yann Gouttenoire and Filippo Sala for fruitful discussions.
The work of R.J. is supported by JSPS KAKENHI Grant Number 24K07013.
Y.N. is supported by Natural Science Foundation of Shanghai.

\appendix

\bibliographystyle{JHEP}
\bibliography{reference}

\end{document}